\documentclass[twocolumn,10pt]{wlscirep}
\usepackage{cuted}

\title{Selective equal spin Andreev reflection at vortex core center in magnetic semiconductor-superconductor heterostructure}

\author[1,2,3]{Chuang Li}
\author[1,2,3]{Lun-Hui Hu}
\author[1,2]{Yi Zhou}
\author[2,3,*]{Fu-Chun Zhang}
\affil[1]{Department of Physics, Zhejiang University, Hangzhou, Zhejiang, 310027, China}
\affil[2]{Kavli Institute for Theoretical Sciences,  University of Chinese Academy of Sciences, Beijing 100190, China}
\affil[3]{Collaborative Innovation Center of Advanced Microstructures, Nanjing, 210093, China}

\affil[*]{fuchun@ucas.ac.cn}

\begin{abstract}
	Sau, Lutchyn, Tewari and Das Sarma (SLTD) proposed a heterostructure consisting of a semiconducting thin film sandwiched between an s-wave superconductor and a magnetic insulator and showed possible Majorana zero mode.  Here we study spin polarization of the vortex core states and spin selective Andreev reflection at the vortex center of the SLTD model. In the topological phase, the differential conductance at the vortex center contributed from the Andreev reflection, is spin selective and has a quantized value $(dI/dV)^{topo}_A =2e^2/h$ at zero bias. In the topological trivial phase, $(dI/dV)^{trivial}_A$  at the lowest quasiparticle energy of the vortex core is spin selective due to the spin-orbit coupling (SOC). Unlike in the topological phase,  $(dI/dV)^{trivial}_A$ is suppressed in the Giaever limit and vanishes exactly at zero bias due to the quantum destruction interference.
	
\end{abstract}

\begin{document}
	\flushbottom
	\maketitle
	\thispagestyle{empty}
	
	\section*{Introduction}
	
	Majorana fermions\cite{Majorana_2008}, whose anti-particles are themselves, were initially envisioned by E. Majorana in elementary particle physics. It has recently been revealed that the Majorana fermions may exist in a number of condensed matter systems\cite{nayak_rmp_2008,elliott_rmp_2015} as zero-energy states so-called Majorana zero modes (MZMs). In the earlier works, chiral p-wave superconductor\cite{sarma_prb_2006} (SC) and $\nu=5/2$ fractional quantum Hall system\cite{more_read_1991} are possible to host MZM in condensed matter systems. In 2008, Fu and Kane\cite{fu_prl_2008} proposed that MZM can be localized in the vortex core by inducing an effective superconducting (SCing) pairing gap on the surface states of a 3-dimensional (3D) strong topological insulator, such as topological insulator (Bi$_2$Te$_3$)/s-wave SC (NbSe$_2$) heterostructure\cite{xu_prl_2015}.  Sau et al.\cite{sau_prl_2010} proposed  a setup by using magnetic insulator/s-wave SC, in which there exists MZM at angular momentum \(m=0\) channel in the vortex core.  Lutchyn et al.\cite{lutchyn_prl_2010} and Oreg et al.\cite{oreg_prl_2010} studied semiconductor Rashba nanowire with strong Rashba SOC and demonstrated localized MZM at the ends of the wire. Mourik\cite{mourik_science_2012} et al. presented evidence for possible existence of non-Abelian MZM in InSb nanowires. Many other theoretical proposals and experimental evidences for MZM have also been reported \cite{Wilczek_np_2009,perge_science_2014,xu_prl_2014,xu_prl_2015,mzm_our_prl_2016,kjaergaard_arxiv_2016}.
	
	To detect the MZMs by transport measurement, the quantized zero-bias peak due to MZM has been theoretically studied\cite{vic_prl_2009}.  He et al. \cite{he_prl_2014} have also proposed Majorana-induced selective equal spin Andreev reflection (SESAR) in 1D Rashba nanowire.   In a usual Andreev reflection\cite{BTK_physics_1982,kashiwaya_rep_pro_2000,tinkham_book_2004} on a topological trivial SC, an incident electron of spin up(down) is reflected with a hole of the opposite spin. MZM is self-conjugate and allows equal spin Andreev reflection. If we assume the MZM is spin-up, then an electron of spin-up is reflected with a hole of the same spin, while an electron of spin-down will have only a normal reflection process being reflected as an electron.
	However, this property of MZM is strongly related to the polarization of the MZM.  In the 2D Fu-Kane model, the polarization of MZM in the vortex core center is controlled by the direction of external magnetic field,  and the tunneling conductance of spin polarization dependence has been observed in experiment\cite{mzm_our_prl_2016,our_prb_2016}, which has provided strong evidence for the existence of MZM.
	
	We note that there is a close similarity between Fu-Kane model and SLTD model. In both models, a topological non-trivial Fermi surface can be realized and an s-wave superconducting pairing can then open a full gap with a chiral MZM at the boundary, and a localized MZM in the vortex core. On the other hand, SLTD model may also give a topological trivial phase in certain parameter space. In this work, we study the 2D semiconductor with SOC hybridization with an s-wave SC.  We will focus on the vortex core states to examine the spin polarization of the MZM as well as other quasiparticle states. In the calculation of the differential tunneling conductance, the spin-polarized scanning tunneling microscope (STM) tip is modeled as a normal lead providing incident particles and receiving scattered particles, as depicted in Fig.~\ref{fig-1-sketch}.
	
	The paper is organized as follows:
	we firstly describe the model Hamiltonian and present the spectra and the corresponding wave functions inside the vortex core in the topological phase. We then apply Fisher-Lee-Landauer-B$\text{\"{u}}$tikker formula to calculate the differential tunneling conductance.  We then discuss the SOC induced SESAR in the topological trivial phase. Finally we will give a brief summary.

\section*{Models \& Results}


\subsection*{Model Hamiltonian for device}\label{sec-Mod_MZM}
\begin{figure}[!htbp]
	\centering
	\includegraphics[width=3.2in]{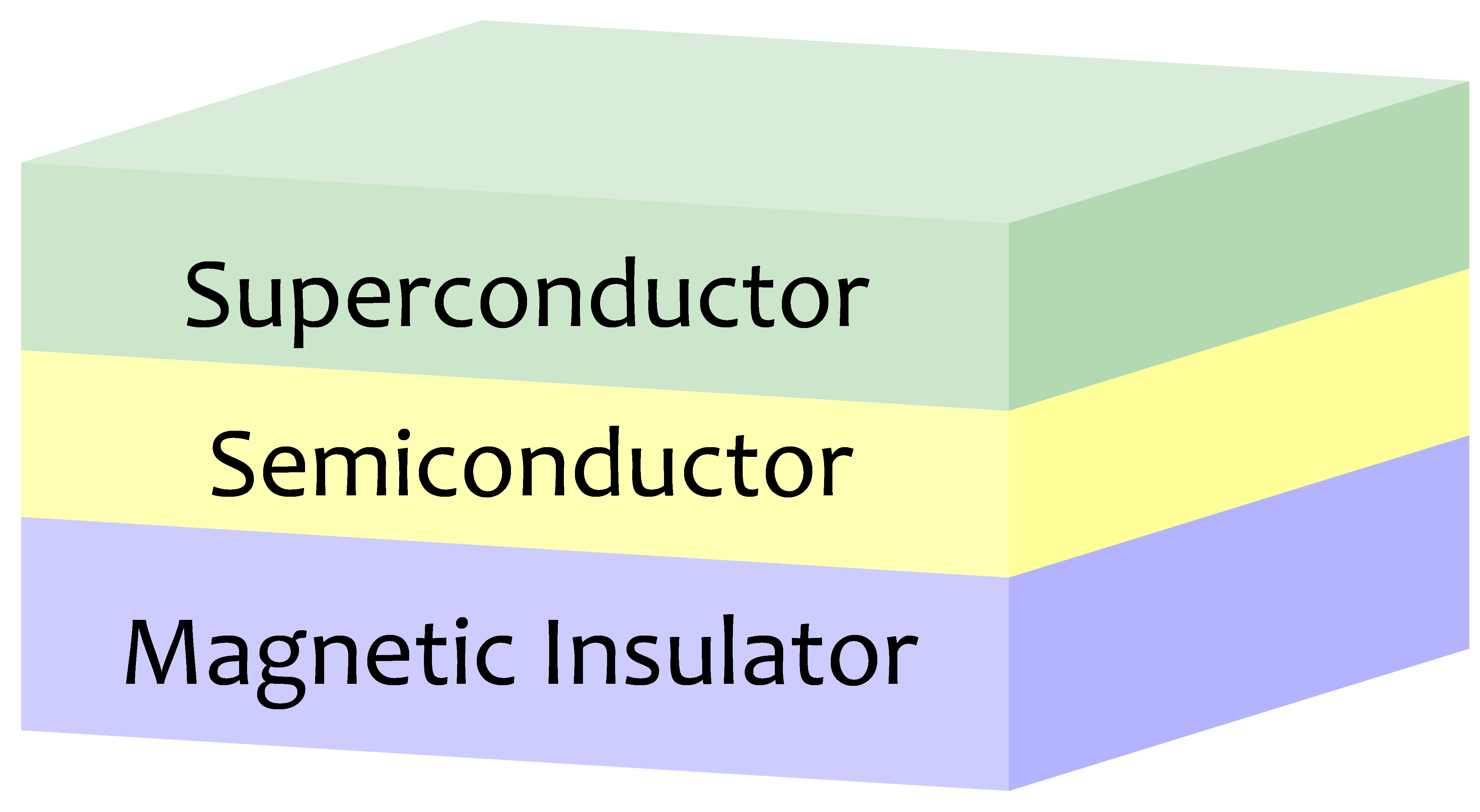}
	\caption{\label{fig-heterostructure}Illustration of the semiconductor-superconductor heterostructure studied in this paper. The semiconductor thin film is described by a 2D electron gas with a Rashba SOC. Superconductivity in the semiconductor is induced by proximity effect. A magnetic field is pointed down which induces a Zeeman energy $V_z$ in Eq.\eqref{eq-h0-part}}
\end{figure}

We study a 2D semiconductor with a Rashba SOC, which is hybridized to an s-wave SC and under a Zeeman field \cite{sau_prl_2010} [see Fig.~\ref{fig-heterostructure}]. The system is described by SLYD model. The Cooper pairs in the semiconductor are induced through the proximity effect. It resembles an effective chiral $p_x+ip_y$ topological SC at the interface. The model Hamiltonian reads,
\begin{align}
	\label{eq-ham-all-continum}
	\mathcal{H}_{\text{D}} &= \mathcal{H}_0 + \mathcal{H}_{\text{SC}} \\
	\label{eq-h0-part}
	\mathcal{H}_{0} &= \int\, d^2\vec{r}\; \tilde{c}^\dagger(\vec{r}) \, \left\lbrack \frac{\hat{p}^2}{2m^\ast} + \alpha_R(\vec{\sigma}\times\vec{p})\cdot\hat{z} - V_z\sigma_z -\mu \right\rbrack \tilde{c}(\vec{r})  \\
	\mathcal{H}_{\text{SC}} &= \int\, d^2\vec{r}\;  \left\lbrack \Delta(\vec{r})\hat{c}_{\uparrow}^\dagger(\vec{r})\hat{c}_{\downarrow}^\dagger(\vec{r}) + \text{H.c.} \right \rbrack
\end{align}
with $m^\ast, \mu, \alpha_R$ and $V_z$ being the effective mass of electron (in the 2D thin film), chemical potential, strength of the Rashba SOC, and the Zeeman field, respectively. The Pauli matrices $\vec{\sigma}=(\hat{\sigma}_x,\hat{\sigma}_y,\hat{\sigma}_z)$ are spin, and the electron annihilation operators read $\tilde{c}(\vec{r}) = \left\lbrack \hat{c}_{\uparrow}(\vec{r}),\,\hat{c}_{\downarrow}(\vec{r}) \right\rbrack^{\text{T}}$. $\Delta(\vec{r})$ is the proximity-induced on-site pairing gap function in the 2D semiconductor. According to the AZ classification\cite{schnyder2008classification}, the above Hamiltonian $\mathcal{H}_{\text{D}}$ belongs to D class since only particle-hole symmetry $K\tau_x$ is preserved, where $K$ is the complex conjugate and $\tau_x$ is an operator describing particle-hole transformation.

The dispersions of the Hamiltonian in Eq.\eqref{eq-h0-part}, corresponding to the helical chirality $\lambda=\pm1$, are,
\begin{align}
E_{\pm} = \frac{\mathbf{k}^2}{2m^\ast} - \mu \pm \sqrt{\alpha_R^2\mathbf{k}^2 + V_z^2}
\end{align}
where we set $\hbar=1$ for convenience. For $V_z \neq 0$ and $|\mu| \le V_z$, there is an energy gap $2V_z$ at $\Gamma$ point ($\mathbf{k}=0$).
If the pairing potential $\Delta(\vec{r})=\Delta_0$ is uniform and if the criterion $V_z^2 > \Delta_0^2 + \mu^2$ is satisfied\cite{oreg_prl_2010,lutchyn_prl_2010,sau_prl_2010,alicea_rep_pro_2012},  the system will open a gap at the dispersion's outer wings without closing the Zeeman gap at $\Gamma$ point. In this case, the system is essentially the same as an effective 2D spinless $p_x+ip_y$ topological SC with chiral MZM\cite{fu_prl_2008}, so we expect a localized MZM in the vortex core of the system.

To study the quasiparticle excitations of the Hamiltonian in Eq.\eqref{eq-ham-all-continum} with a single vortex, we use Bogoliubov-de Gennes (BdG) equation,
\begin{align}\label{eq-bdg}
\left(
\begin{array}{cc}
\mathcal{H}_{0} & -i\sigma_y \Delta(\vec{r}) \\
i\sigma_y\Delta(\vec{r}) & -\mathcal{H}_0^\ast \\
\end{array}
\right) \Psi_n(\vec{r}) = E_n \Psi_n(\vec{r})
\end{align}
where the Nambu spinor notation $\Psi(\vec{r}) = \left\lbrack u_\uparrow(\vec{r}),\,u_{\downarrow}(\vec{r}),\,v_{\uparrow}(\vec{r}),\,v_{\downarrow}(\vec{r}) \right\rbrack^{\text{T}}$ is used here. Because of the particle-hole symmetry in this BdG equation \eqref{eq-bdg}, both $\left\lbrack u_\uparrow(\vec{r}),\,u_{\downarrow}(\vec{r}),\,v_{\uparrow}(\vec{r}),\,v_{\downarrow}(\vec{r}) \right\rbrack$ and $\left\lbrack -v_\uparrow^\ast(\vec{r}),\,-v_{\downarrow}^\ast(\vec{r}), \,u_{\uparrow}^\ast(\vec{r}), \,u_{\downarrow}^\ast(\vec{r}) \right\rbrack$ are eigenfunctions with eigenenergies $E_n$ and $-E_n$, respectively.  The Bogoliubov quasiparticle operator is defined as,
\begin{align}\label{eq-gamma-gammadagger}
\gamma_n^\dagger = \int\,d\vec{r}\; \sum_{s} \left\lbrack  u_{n s}(\vec{r}) \hat{c}^{\dagger}_{s}(\vec{r}) + v_{n s}(\vec{r})\hat{c}_{s}(\vec{r}) \right\rbrack
\end{align}
The necessary condition for MZM is $\gamma^\dagger =\gamma$ for zero-energy mode.

\begin{figure}[!htbp]
	\centering
	\includegraphics[width=3.2in]{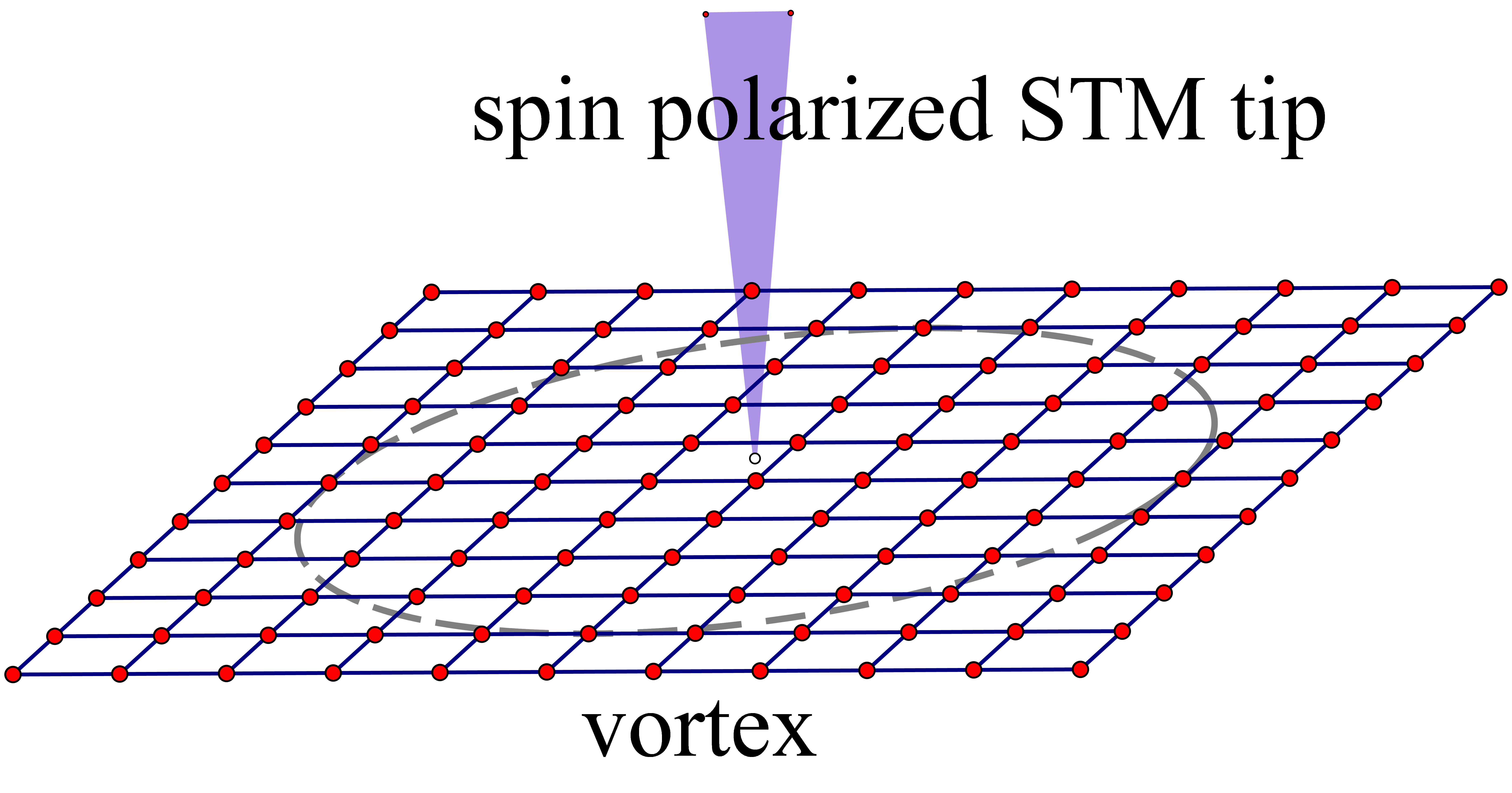}
	\caption{\label{fig-1-sketch}Illustration of the system.}
\end{figure}

There are three main practical approaches to solve BdG equation \eqref{eq-bdg}. The first one is to use corresponding  2D tight-binding model\cite{liang_epl_2012,zhou_epl_2013,li_sci-rep_2014} of Hamiltonian in Eq.\eqref{eq-ham-all-continum}, and solve the problem in a lattice.  The second one is to adopt a disc geometry to solve the BdG equation \eqref{eq-bdg} and use orthogonal Bessel functions\cite{gygi_prb_1991,hayashi_prl_1998,sau_prl_2010,mao_prb_2010,huo_prl_2012,kawakami_prl_2015}. And the third one is to adopt spherical geometry to utilize harmonic spherical function or associated Legendre polynomials\cite{kraus_prl_2008,kraus_prb_2009,sau_prb_2010,mzm_our_prl_2016}. In this work, we shall use tight-binding model on 2D square lattice [see Fig.~\ref{fig-1-sketch}], where the Hamiltonian becomes,
\begin{align}
	\label{eq-ham-all-lattice}
	\mathcal{H}_{\text{D}} &= \mathcal{H}_0 + \mathcal{H}_{\text{SC}} \\
	\mathcal{H}_0  &= -t_D \sum_{\langle ij \rangle,s}(c^{\dag}_{i s} c_{j s}) +(4t_D-\mu_D)\sum_{i,s}(c^{\dag}_{i s} c_{i s}) \nonumber \\
	&\quad -i \frac{\alpha_R}{2} \sum_{\langle ij \rangle,s, s'}(\boldsymbol{\sigma}_{s s'} \times \hat{\boldsymbol{r}}_{ij})_z
	c^{\dag}_{i s} c_{j s'}  \\
	&\quad - V_{z,D} \sum_{i,s} \sigma_z
	c^{\dag}_{i s} c_{i s'} \nonumber \\
	\mathcal{H}_{\text{SC}} &= - \sum_{i} \left\lbrack \Delta_i c^{\dag}_{i \uparrow} c^{\dag}_{i \downarrow} + \text{H.c.} \right\rbrack
\end{align}
where \( t_D \) and \( \mu_D \) are the nearest-neighboring hopping integral and the chemical potential, respectively, \( \alpha_R \) is the Rashba SOC strength, and \( \hat{\boldsymbol{r}}_{ij} \) denotes the unit vector between site \( i \) and \( j \). \(V_{z,D}\) is the Zeeman energy due to the z-direction magnetic field. We assume that the pairing order parameter inside the vortex has form
\begin{align}\label{eq-vortex}
\Delta (\vec{r})= \Delta_0 \text{tanh}(r/ \xi) e^{i \varphi}
\end{align}
where \(r\) is the distance of the lattice site from the vortex core, and \( \varphi \) is the azimuthal angle of \( \vec{r} \), and \( \xi \) describes the size of the vortex. We diagonalize the BdG equation \eqref{eq-bdg} for Hamiltonian in Eq.\eqref{eq-ham-all-lattice} to obtain the eigenvalues and the corresponding eigenvectors  by using the Feast Eigenvalue Solver for large sparse matrix. In this paper, we consider a lattice size of \( N= 199 \times 199 \) with open boundary condition, and the vortex is located at the center of the lattice. The Hamiltonian in Eq.\eqref{eq-bdg} is a $4N\times4N$ hermitian matrix.

\begin{figure}[!htbp]
	\centering
	\includegraphics[width=3.2in]{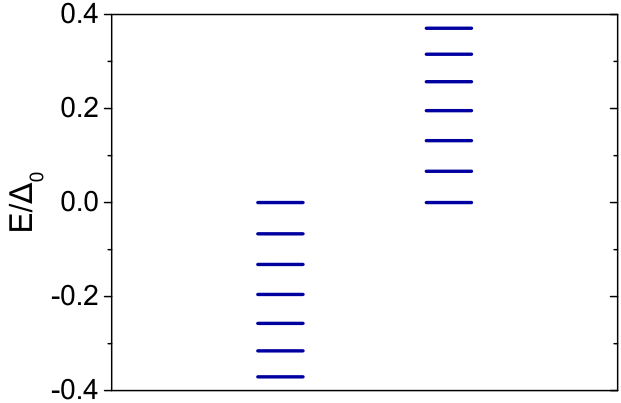}
	\caption{\label{fig-2-EigenEnergy} Low lying eigenenergies of the vortex states on 2D square lattice of the Hamiltonian.  The parameters are: \( t_D= 1.0 \), \( \mu_D= 0 \), \( \alpha_R= 1.8 \), \( V_{z,D}= 0.8\), \( \Delta_0= 0.5 \), and \( \xi= 8.0 \) in Eq.\eqref{eq-vortex}.}
\end{figure}

\subsection*{Spin polarized MZM in the Vortex Core}\label{sec-Result_MZM}
\begin{figure}[!htbp]
	\centering
	\includegraphics[width=3.2in]{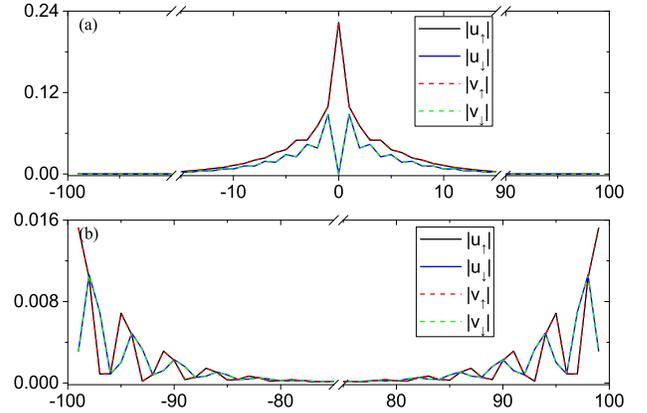}
	\caption{\label{fig-3-WaveFunc_MF} Amplitude of the MZMs wave-functions (a) in the vortex core; and (b) at the edge. The horizontal axis indicates the lattice site along (0,1) direction with the vortex core centered at (0,0), the center of the square lattice.  The wave-functions of electron and hole components are identical for each spin, consistent with the requirement for Majorana fermion.}
\end{figure}

The energy spectra of the vortex states are plotted in Fig.~\ref{fig-2-EigenEnergy} in topological phase region with parameters given in the figure caption.  In the topological phase, we expect a MZM in the vortex core and a MZM at the edge in the infinitely large system.  In a finite size system, the vortex core state and the edge state have a hybridization, leading to a pair of the MZMs with energies $\pm E_0$ very close to zero.  Our numerical calculations agree with this analysis and the calculated  $E_0 \sim 10^{-6}$.  By linear recombination of the two MZMs, we find a MZM localized in the vortex core, [see Fig.~\ref{fig-3-WaveFunc_MF}(a)], and the other one is localized at the edge [see Fig.~\ref{fig-3-WaveFunc_MF}(b)]. Both satisfy the MZM condition $\gamma^\dagger = \gamma$ in Eq.\eqref{eq-gamma-gammadagger} to a high accuracy. Note that we only consider a single vortex in our model calculation, since the edge MZM will be easily destroyed in real system, where there are many vortices. Below we will only focus on the bound states in the vortex core. As shown in Fig.~\ref{fig-3-WaveFunc_MF}(a), one can see that the MZM's wave-function at the center of the vortex is fully polarized with spin-up: $\vert u_\uparrow\vert\neq0$ and $\vert u_\downarrow\vert = 0$. As a comparison, the wave function of the first excited state is shown in Appendix A, which is spin-down at the center of the vortex.

\subsection*{Transport calculation}
To experimentally detect the  MZMs, it is important to examine some unique transport features of the MZMs, such as quantized conductance \cite{vic_prl_2009} and SESAR \cite{he_prl_2014,mzm_our_prl_2016,our_prb_2016}.
In this work, we theoretically study  spin-polarized transport properties of the MZM in the vortex core in the model, which can be tested in STM/STS measurement. We consider the STM tip as a 1D normal lead, illustrated in Fig.~\ref{fig-1-sketch}, and the Hamiltonian $\mathcal{H}_{L}$ for the semi-infinity lead is,
\begin{align}\label{eq-Hmtn-Ld}
\begin{split}
	\mathcal{H}_L &= -t_L \sum_{\langle ij \rangle,s}(c^{\dag}_{i s} c_{j s}) +(2t_L-\mu_L)\sum_{i,s}(c^{\dag}_{i s} c_{i s}) \\
	&\quad - \sum_{i,s,s'} (\boldsymbol{V}_L \cdot \boldsymbol{\sigma}_{s s'})(c^{\dag}_{i s} c_{i s'})
\end{split}
\end{align}
here \( t_L \) and \( \mu_L \) are lead's nearest-neighboring hopping coefficient and chemical potential respectively. \( \boldsymbol{V}_L \) is the potential of the magnetic field on the lead.
The lead's vertex (site \( 0 \)) contacts the device at site \( p \), then
\begin{equation}\label{eq-Hmtn-Cp}
\mathcal{H}_c = -t_c \sum_{s} \left\lbrack c^{\dag}_{L 0 s} c_{D p s} + \text{H.c.} \right\rbrack
\end{equation}
where the lattice labels $L0$ and $Dp$ are the connected (touched) points from N/S junction. Here we use \( t_L= 1.2 \), \( \mu_L= 0 \) for the lead, and choose \( |\boldsymbol{V}_L|= 0.4 \) to polarize the spin, adjusting the spin polarized direction to the local wave function's of the vortex MZM. And, we set the coupling coefficient \( t_c= 0.6 \) in the Giaever limit, which can simulate the barrier strength at the interface between normal lead and the SC device.

Now we consider a single electron with energy \( E \) injected from the spin polarized normal lead (STM tip, shown in Fig.~\ref{fig-1-sketch}), and reflected by the SLTD device. We use the basis \( \left\lbrack\begin{array}{cccc}
c_{i \uparrow} & c_{i \downarrow} & c^{\dag}_{i \uparrow} & c^{\dag}_{i \downarrow}
\end{array}\right\rbrack^\text{T} \) in our calculation.
In this situation, the \( 4 \times 4 \) submatrix of scattering matrix (S-matrix) at the probed point site \( p \),
\begin{align}
S(p,p)= \left( \begin{array}{cc}
r_{ee} & r_{eh} \\
r_{he} & r_{hh}
\end{array} \right)
\end{align}
describes the property of SESAR.
$ \text{Tr}( r_{ee}^{\dag} r_{ee} + r_{he}^{\dag} r_{he} )=2 $, because of unitary of S-matrix.

Base on the S-matrix, the differential tunneling conductance contributed from the Andreev reflection can be calculated by using Landauer-B\( \ddot{\text{u}} \)tikker formula\cite{landauer_mag_1970,buttiker_prb_1988,meir_prl_1992,datta_book_1997}
\begin{align}\label{eq-S-matrix-didv}
\begin{split}
dI/dV & = 2 - \text{Tr}(r^{\dag}_{ee} r_{ee}) + \text{Tr}(r^{\dag}_{he} r_{he}) \\
& = 2 \text{Tr}(r^{\dag}_{he} r_{he})
\end{split}
\end{align}
where we have used \( e^2/h \) for the unit of the conductance.

\subsection*{SESAR in topological phase}\label{sec-Result_SESAR}
\begin{figure}[!htbp]
	\centering
	\includegraphics[width=3.2in]{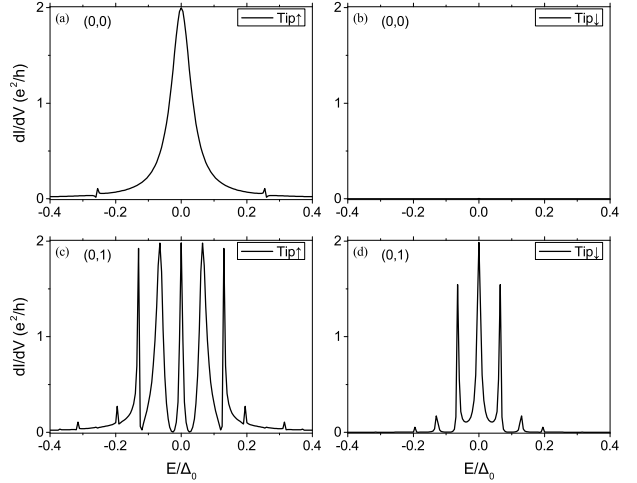}
	\caption{\label{fig-5-SESAR} Local differential tunneling conductance contributed from the Andreev reflection as function of incident electron's energy at the vortex core (0,0) for the STM tip spin-up (a) and spin-down (b), which is zero, and at site (0,1) for the STM tip spin-up (c) and spin-down (d).  Parameters in the calculations are the same as in Fig.~\ref{fig-2-EigenEnergy}.}
\end{figure}

We calculate the differential tunneling conductance \( dI/dV(E)\) as a function of incident electron's energy in Eq.\eqref{eq-S-matrix-didv}. Usually as for the STM experiment\cite{mzm_our_prl_2016}, the measured conductance comes from two parts: normal conductance (proportional to local density of states) and Andreev reflection\cite{sun_prb_1999,wang_prl_2009,yang_prl_2010,huang_epl_2012}. In this work, we only focus on the Andreev reflection part. Andreev reflection due to MZM is very different from that in usual SC.  In the usual SC case, Andreev reflection is known to be weak and can be neglected in the Giaever limit. But Andreev reflection due to MZM is very different and its strength at zero energy remains $2e^2/h$ as pointed out previously and also shown in our numerical results below in Fig.~\ref{fig-5-SESAR}.

The differential conductance $dI/dV$ as functions of energy due to the Andreev reflection are plotted in Fig.~\ref{fig-5-SESAR} for a topological phase of the model.  dI/dV is spin selective and shows a quantized zero-bias peak, i.e., $dI/dV = 2 \,e^2/h$ at the center of the vortex core.  We can analyse the S-matrix in Eq.\eqref{FL eq.}, and see that the outgoing hole is spin up for spin-up incident electron. This is the reason why spin polarized STM experiment can see the unique signal of MZM. As we expect, the width of the zero-bias peak at the lattice site away from the core center becomes narrow and the spin polarization dependence becomes weak.

\subsection*{SOC induced SESAR in topological trivial phase}
\begin{figure*}[!htbp]
	\centering
	\includegraphics[width=6.4in]{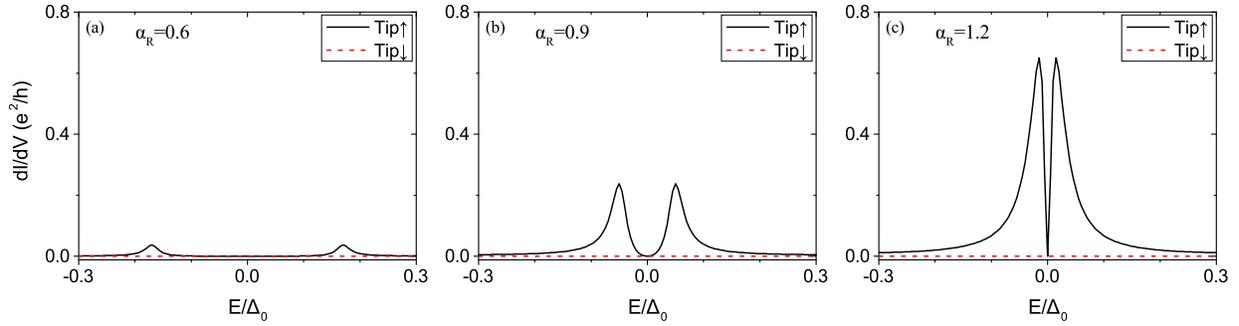}
	\caption{\label{fig-5-SESAR-triv-soc} Local differential tunneling conductance at the vortex core center of the model in topologically trivial phase \( V_{z,D}= 0 < \sqrt{\Delta^2+\mu_D^2} \), for various values of SOC strength $\alpha_R$.  Other parameters in the calculations are the same as those in Fig.~\ref{fig-2-EigenEnergy}.}
\end{figure*}

\begin{figure*}[!htbp]
	\centering
	\includegraphics[width=6.4in]{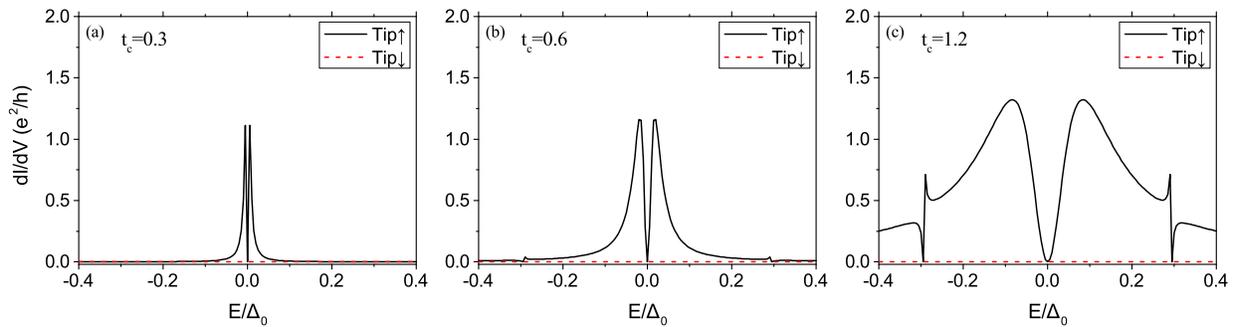}
	\caption{\label{fig-5-SESAR-triv} Local differential tunneling conductance at the vortex core center of the model in topologically trivial phase \( V_{z,D}= 0 < \sqrt{\Delta^2+\mu_D^2} \), for various values of the coupling $t_c$ between the device and the lead.  Other parameters in the calculations are the same as those in Fig.~\ref{fig-2-EigenEnergy}.}
\end{figure*}

In this subsection, we study the topological trivial phase. We use same models Eq.\eqref{eq-ham-all-lattice}\eqref{eq-Hmtn-Ld}\eqref{eq-Hmtn-Cp} and methods in previous sections, and keep parameters unchanged except setting, for simplicity, $V_{z,D}=0<\sqrt{\Delta^2+\mu_D^2}$ which belongs to topological trivial region. 
We perform numerically calculation for the spin-polarized differential conductance, and the results are shown in Fig.~\ref{fig-5-SESAR-triv-soc} and Fig.~\ref{fig-5-SESAR-triv}. 

Interestingly, we find there is no SESAR signal at \( E=0 \), namely $dI/dV(0)=0$ exactly, as  Fig.~\ref{fig-5-SESAR-triv-soc} and Fig.~\ref{fig-5-SESAR-triv}. It indicates perfect quantum destructive interference\cite{yamakage_pe_2014,ii_prb_2012}. And it can be explained by the particle-hole symmetry which makes the anomalous Green's function\cite{our_prb_2016} vanishing in the vortex core center,
\begin{align}
G_0^R(r=0,E) = \sum_n \frac{\vert\Psi_n(r=0)\rangle \langle \Psi_n(r=0)\vert }{E-E_n + i\delta} \to 0
\end{align}
because of $\vert E_n\vert\gg\delta$. However, it will not happen if there exists MZM whose energy is almost zero.

Besides, in topological trivial case, for ground state wave function, the electronic component $|u_\uparrow|$ is no longer equal to the hole part $|v_\uparrow|$. Since electron-hole reflection $r_{he}$ by spin-up particles $\sim |u_\uparrow v_\uparrow|$ while $r_{ee} \sim |u_\uparrow^2-v_\uparrow^2|$, maximum of the differential conductance $dI/dV$ is less than $\frac{2e^2}{h}$.

When the SOC strength is increasing, as in Fig.~\ref{fig-5-SESAR-triv-soc}, the SESAR signal becomes stronger and stronger. And it leads to our main conclusion that SESAR can be induced by SOC. This signal is different from the MZM-based SESAR, since there is no MZM in topological trivial phase.
In Fig.~\ref{fig-5-SESAR-triv}, we vary the hopping element $t_c$ from $0.3$ (dirty limit) to $1.2$ (transparent limit), the conductance becomes more and more broad, consistent with the BTK theory\cite{BTK_physics_1982}.
In a short conclusion, the condition for the appearance of SOC-induced SESAR is large SOC strength and small tunneling barrier between STM tip and device.

To analytically investigate the effect of the SOC on the localized vortex core states, we start with a discussion of the Hamiltonian in continuum  space on a spherical surface \cite{our_prb_2016}. Note that the finite size effect can be reduced by increasing the radius of sphere. The Hamiltonian reads,
\begin{align}
\begin{split}
	\mathcal{H}'_0 &= \left(\frac{\eta}{R^2}L^2 - \frac{\alpha_R}{R}\vec{L}\cdot\vec{\sigma}-\mu_D \right)\otimes \tau_z \\
	&\quad + \Delta(\theta) I\otimes \left(\cos(\phi)\tau_x + \sin(\phi)\tau_y  \right)
\end{split}
\end{align}
where the Nambu basis is $\{c_{\uparrow}, c_{\downarrow}, c_{\downarrow}^\dagger, -c_{\uparrow}^\dagger\}$. The SC pairing function is $\Delta(\theta)=\Delta_0 \tanh(R\sin(\theta)/\xi)$ with $R$ as radius of sphere, and $\theta$ for polar angle, $\phi$ for azimuth angle. We firstly turn off SOC, i.e., $\alpha_R=0$, and then turn on SOC as perturbation term. In the absence of SOC ($\alpha_R=0$), the 4-by-4 BdG Hamiltonian can be decoupled into two 2-by-2 blocks. In each block, we notice that $L_z+\sigma_z/2$ or $L_z-\sigma_z/2$ provide a good quantum number, thus $\vert m,i\rangle$ ($i=\pm$ represents blocks) can be used to label the in-gap vortex states: $(L_z\pm\sigma_z/2)\vert m,\pm\rangle=(m\pm1/2)\vert m,\pm\rangle$. Therefore, the wave function for the localized vortex core states are $\vert m,+\rangle = \left\lbrack e^{im\phi} u_{\uparrow,m}(\theta), 0 , e^{i(m+1)\phi}v_{\downarrow,m}(\theta),0 \right\rbrack$ and $\vert m,-\rangle = \left\lbrack 0, e^{im\phi} u_{\downarrow,m}(\theta),0 , e^{i(m+1)\phi}v_{\uparrow,m}(\theta) \right\rbrack$, both satisfying $\mathcal{H}_0\vert m, \pm \rangle = E_m \vert m,\pm \rangle$. The components for the corresponding wavefunction are,
\begin{align}
	u_{\uparrow,m}(\theta) &= u_{\downarrow,m}(\theta) = \sum_{l\ge\vert m\vert} A_{l}^m \, Y_{l}^{m}(\theta)  \\
	v_{\uparrow,m}(\theta) &= v_{\downarrow,m}(\theta) = \sum_{l\ge\vert m\vert} B_{l}^m \, Y_{l}^{m+1}(\theta)
\end{align}
where $A_l^m$ and $B_l^m$ are the corresponding coefficients, and $Y_l^m(\theta)=P_l^m(\cos\theta)/\sqrt{2\pi}$ with $P_l^m$ the associated Legendre polynomial. Note that the $m=0$ channel is in the particle-hole relationship with the $m=-1$ channel, the $m=1$ channel is in the particle-hole relationship with the $m=-2$ channel. And we also assume that the quasiparticle wave functions are all orthogonal and normalized.

Then, we switch on the SOC ($\alpha_R\neq0$), and assume it is small compared with $\eta$. For the following discussion, we only focus on the vortex core center. In the presence of SOC, $K_z=L_z+(\sigma_z-\tau_z)/2$ provides a good quantum number\cite{our_prb_2016}. Then, the SOC will mix the in-gap states, so that the corrected wave function should be eigenvector of $K_z$. And we find that the corrected wave function up to the first order is exact since all spherical harmonics with $l\ge1$ vanish in the vortex core center. In other words, only $u_{\uparrow/\downarrow,0}$ and $v_{\uparrow/\downarrow,0}$ survives. Thus the corrected wave function is given by
\begin{align}
    \widetilde{\vert m,i\rangle} = \vert m,i\rangle + \lambda \sum_{n\neq m,j=\pm}  \vert n,j\rangle \frac{\langle n,j \vert \mathcal{H}'_\lambda \vert m,i\rangle}{E_m - E_n}
\end{align}
where we denote $\mathcal{H}'_\lambda=\lambda\vec{L}\cdot\vec{\sigma}\otimes\tau_z$ with $\lambda=-\frac{\alpha_R}{R}$. Thus the corrected wave function for $\vert 0,\pm\rangle$ and $\vert -1,\pm\rangle$ are,
\begin{align}
  	\widetilde{\vert 0,+\rangle}  &= \sqrt{1-\vert A\vert^2} \vert 0,+\rangle + A \vert -1,-\rangle \\
  	\widetilde{\vert -1,-\rangle} &= \sqrt{1-\vert A\vert^2} \vert -1,-\rangle - A^\ast \vert 0,+\rangle \\
  	\widetilde{\vert 0,-\rangle}  &= \sqrt{1-\vert B\vert^2} \vert 0,-\rangle + B \vert 1,+\rangle \\
  	\widetilde{\vert -1,+\rangle} &= \sqrt{1-\vert C\vert^2} \vert -1,+\rangle + C \vert -2,-\rangle
\end{align}
where $A,B,C$ are normalized coefficients and proportional to $\lambda = -\frac{\alpha_R}{R}$. Then we observe that both $\widetilde{\vert 0,+\rangle}$ and $\widetilde{\vert -1,-\rangle}$ have nonzero electron component $u_{\uparrow}$ and hole component $v_{\uparrow}$, which are spin polarized in the vortex core center. Furthermore, the equal spin anomalous Green's function (proportional to $u_{\uparrow}v_{\uparrow}^\ast$ or $u_{\downarrow}v_{\downarrow}^\ast$) contributes to the Andreev reflection. From this point of view, we may expect that there is also equal spin Andreev reflection in the vortex core center, even when the system is topological trivial. However, as for $\widetilde{\vert 0,-\rangle}$, $\widetilde{\vert -1,+\rangle}$, $\widetilde{\vert 1,+\rangle}$ and $\widetilde{\vert -2,-\rangle}$ contain only nonzero electron component $u_\downarrow$ or hole component $v_\downarrow$. They are also spin polarized down, but there is no Andreev reflection signal. Therefore, this phenomena can also be treated as SESAR.

Moreover, we should emphasize that such SESAR is totally induced by SOC, if the SOC strength is larger, $uv^\ast\sim\alpha_R$ will be larger, then the SESAR effect become more and more obvious, which is consistent with numerical results of square lattice in Fig.~\ref{fig-5-SESAR-triv-soc}.

\section*{Summary}\label{sec-Con}
In this work, we have studied selective equal spin Andreev reflection at vortex core center in magnetic semiconductor-superconductor heterostructure described by the model proposed by Sau, Lutchyn, Tewari and Das Sarma. We solve the BdG equation for a single vortex of the model in 2D square lattice. In the topological phase, the Majorana zero mode is localized at the vortex core and its spin component at the center is completely parallel to the external magnetic field, which leads to spin selective Andreev reflection. In the topological trivial phase, there is no Majorana zero mode inside the vortex. However, the spin-orbit coupling induces a spin selective Andreev reflection at the bias of the lowest quasiparticle energy. The Majorana zero mode induced spin selective Andreev reflection is robust and gives a quantized value of differential conductance  $2e^2/h$, which is independent of the tunneling barrier. This quantized spin selective Andreev reflection is consistent with previous theoretical study \cite{our_prb_2016}. The usual vortex quasiparticle induced spin selective Andreev reflection gives a vanishing value of the differential conductance at zero bias due to quantum destructive interference and is sensitive to the barrier in the tunneling.

\section*{Methods}
\subsection*{Transport Methods for $dI/dV$ of N/S Junction}\label{sec-Method_Transport}
Then, to calculate the differential conductance in Eq.\eqref{eq-S-matrix-didv}, what we need is the S-matrix for N/S junction. From the Fisher-Lee Formula\cite{fish-lee_prb_1981}, the whole \( 4N \times 4N \) S-matrix can be calculated
\begin{equation}\label{FL eq.}
S = -1 + i \Gamma^{1/2} G^R \Gamma^{1/2}
\end{equation}
here \( \Gamma= i [ \Sigma^R - ( \Sigma^R )^{\dag} ] \), \( \Sigma^R \) is the self-energy induced by the lead. It's a \( 4N \times 4N \) matrix but only nonzero for the probed site $p$ on the device, that we will show later, as well as \( \Gamma \). And the total (retarded) Green's function reads
\begin{equation}\label{GFunc of device}
G^R = [ E - \mathcal{H}_D - \Sigma^R + i \eta_D ]^{-1}
\end{equation}
Only the \( 4 \times 4 \) sub-matrix \( G^R(p,p) \) contributes to the calculation because of the sparsity of \( \Gamma \). Infinitesimal positive number \( \eta_D=10^{-5} \) is adopted. To calculate the inverse of large sparse matrix, we use Intel MKL PARDISO Solver in Fortran code.

As we know, the lead will contribute a self-energy\cite{datta_book_1997} to the device, which could be described by a \( 4N \times 4N \) matrix \( \Sigma^R \),
\begin{align}
\Sigma^R = \tau^{\dag} G^R_L \tau
\end{align}
\( G^R_L \) is the retarded Green's function of the lead. The coupling matrix \( \tau \) is nonzero only between the adjacent points of the lead's vertex $(L)0$ and device's probed site $(D)p$, \( \tau (L0,Dp) = \text{diag}\{ t_c, t_c, -t_c, -t_c \} \). So the self-energy \( \Sigma^R \) only has a nonzero \( 4 \times 4 \) sub-matrix at the probed point site \( p \) (connected point for N/S junction).
For definition for the surface Green's function of lead,
\begin{equation}\label{eq-GFunc of lead}
[ E - \mathcal{H}_L + i \eta_L ] G^R_L  =1
\end{equation}
\( \eta_L \) is a infinitesimal positive number, we used \( \eta_L=10^{-5} \). Although the lead's Hamiltonian has infinite dimension in matrix form, same as the Green's function, the only submatrix we need in the calculation is the Green's function on the lead's vertex site \( 0 \), because of the sparsity of the coupling matrix \( \tau \). And the surface Green's function \( G^R_L(0,0) \) can be calculated by decimation method through Eq. \eqref{eq-GFunc of lead}.

In order to get the surface Green's function, let's write Eq.\eqref{eq-GFunc of lead} in block matrix form
\begin{align}\label{GFunc of lead _mtx}
\left( \begin{array}{cccc}
d & -A & & \\
-B & D & -A & \\
& -B & D & \ddots \\
& & \ddots & \ddots \\
\end{array} \right)
\left( \begin{array}{cccc}
g_{11} & g_{12} & g_{13} & \cdots \\
g_{21} & g_{22} & g_{23} & \\
g_{31} & g_{32} & g_{33} & \\
\vdots & & & \ddots \\
\end{array} \right) = 1
\end{align}
each letter presents a \( 4 \times 4 \) sub-matrix with different site indexes. Initially, \( d= D \), \( B= A^{\dag} \). In our case, from Eq.\eqref{eq-GFunc of lead}, we have
\begin{strip}
	\begin{align}
	D= \begin{pmatrix}
	E+i\eta -(2t_L - \mu_L - V_z) & +(V_x - i V_y) & & \\
	+(V_x + i V_y) & E+i\eta -(2t_L - \mu_L + V_z) & & \\
	& & E+i\eta +(2t_L - \mu_L - V_z) & -(V_x + i V_y) \\
	& & -(V_x - i V_y) & E+i\eta +(2t_L - \mu_L + V_z) \\
	\end{pmatrix}
	\end{align}
\end{strip}
and the effective interaction between adjacent sites $ A= \text{diag}\{ -t_L,-t_L,t_L,t_L \} $. Note that the only value that we need is \( g_{11} = G^R_L(0,0) \), for the surface Green's function of the lead. We separate Eq.\eqref{GFunc of lead _mtx} into equations in the following forms,
\begin{align}\label{decimation2}
-B g_{n-1,m} +D g_{n,m} -A g_{n+1,m} = \delta_{n,m}
\end{align}
for \( n= 2,3,4,\dots \), and \( m= 1,2,3,\dots \). Write Eq.\eqref{decimation2} with adjacent indices then cancel the two terms: \( g_{n-1,m} \) and \( g_{n+1,m} \), we finally get:
\begin{align}\label{eq-decimation3}
\begin{split}
&-B D^{-1} B g_{n-2,m}-A D^{-1} A g_{n+2,m} \\
&+( -B D^{-1} A +D -A D^{-1} B )g_{n,m} \\
=&\,\delta_{n-1,m} B D^{-1} +\delta_{n,m} +\delta_{n+1,m} A D^{-1}
\end{split}
\end{align}

If we discard all even number for \( n \) and \( m \), and take the transform
\begin{equation} \label{eq-steps-decimation}
\left\{ \begin{array}{l}
A'= A D^{-1} A \\
B'= B D^{-1} B \\
D'= D -B D^{-1} A -A D^{-1} B
\end{array} \right.
\end{equation}
then Eq.\eqref{eq-decimation3} becomes
\begin{equation}
-B' g_{n-2,m} + D' g_{n,m} -A' g_{n+2,m} = \delta_{n,m}
\end{equation}
which has the same form as Eq.\eqref{decimation2}, with $n=3,5,7,\dots$, and $m=1,3,5,\dots$.

Similarly, we have the transform $ d'= d -A D^{-1} B $ for the surface. Then Eq.\eqref{GFunc of lead _mtx} becomes
\begin{align}
\left( \begin{array}{cccc}
d' & -A' & & \\
-B' & D' & -A' & \\
& -B' & D' & \ddots \\
& & \ddots & \ddots \\
\end{array} \right)
\left( \begin{array}{cccc}
g_{11} & g_{13} & g_{15} & \cdots \\
g_{31} & g_{33} & g_{35} & \\
g_{51} & g_{53} & g_{55} & \\
\vdots & & & \ddots \\
\end{array} \right)=1
\end{align}
Repeat these steps in Eq.\eqref{eq-steps-decimation}, we will update the coefficient and abandon the Green's function between nearby sites consistently.

After sufficient number of iterations, the coefficient \( A \) becomes the effective interaction between pretty far sites which must be a sub-matrix comprised of small values. From \( d'' g_{11} -A'' g''_{21}= 1 \), we finally obtain the surface Green's function for lead
\begin{equation}
g_{11}= d''^{-1}
\end{equation}
where $g_{11}$ is exactly the surface Green's function defined in Eq.\eqref{eq-GFunc of lead}.

\subsection*{Apply sparse matrix to Transport calculation}
In the calculation of Eq.\eqref{GFunc of device}, it's difficult to take the inverse directly, since Hamiltonian of SLTD model device  $\mathcal{H}_D$ in dominator is a big matrix of size $4N \times 4N$. Traditionally, the methods of recursive Green's function could be applied here to calculate the Green's function on the contact point $p$, $G^R (p,p)$, for S-matrix calculating Eq.\eqref{FL eq.}.
There is an alternative approach using sparse matrix. In this regard, we store $\mathcal{H}_D$ as sparse matrix, which is suitable for tight-binding lattice model. Then the $j$-th column of total Green's function, defined as $G^R_j$, can be solved by the linear equation
\begin{align}\label{eq-GFunc-Col}
	[ E - \mathcal{H}_D - \Sigma^R + i \eta ] G^R_j =
	\begin{pmatrix}
	0 \\ \vdots \\ 0 \\ 1 (j\text{ row}) \\ 0 \\ \vdots \\ 0
	\end{pmatrix}
\end{align}
and $G^R (p,p)$ can be drawn from relevant columns of total Green's function.

Intel MKL PARDISO Solver could be used to solve the linear equation of sparse matrix Eq.\eqref{eq-GFunc-Col}. We also tried methods of recursive Green's function, by calculating Green's function of one lattice's column gradually. Same result of the differential conductance [Fig.~\ref{fig-5-SESAR}(a)] was obtained, but it cost more time in calculation. So taking inverse of sparse matrix directly shows its efficiency advantage.

\bibliography{reference}

\section*{Acknowledgments (not compulsory)}

We acknowledge helpful discussions with Chih-Chieh Chen, Chui-Zhen Chen, Fei Ye, Jin-Hua Gao, Wei-Qiang Chen, and K. T. Law. Especially, we benefit a lot from detailed communications with Jin-Feng Jia, Haohua Sun,
and James Jun He. 
The work is supported by National Key Research and Development Program of China (No.2016YFA0300202),
National Basic Research Program of China (No.2014CB921201/2014CB921203),
NSFC (No.11374256/11674278/11774306) and the Fundamental Research Funds for the Central Universities in China.

\section*{Author contributions statement}
C.L. performed the calculations with assistance from L.H.H., and provided all of the figures. Y.Z. and F.C.Z. initiated and supervised the project. All authors analysed the results and wrote the paper.

{\bf Corresponding author:} Correspondence to Fu-Chun Zhang.

\section*{Additional information}
{\bf Competing financial interests:} The authors declare no competing financial interests.

\clearpage
\begin{center}
\section*{Appendix}
\end{center}

\subsection*{Appendix A : Low-Energy states}
\begin{figure}[!htbp]
  	\centering
  	\includegraphics[width=3.2in]{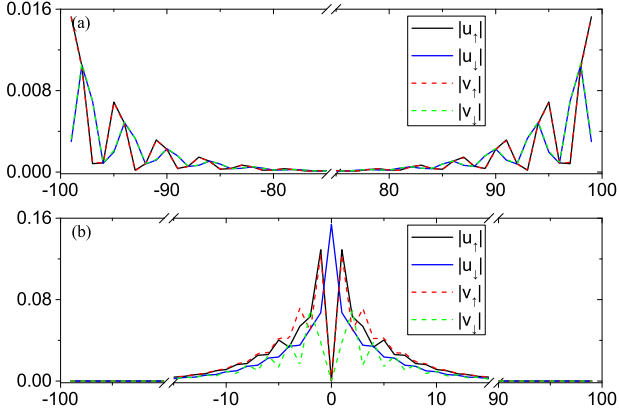}
  	\caption{\label{fig-4-WaveFunc-excitation} The first excitation for a) edge state; b) vortex state.}
  \end{figure}

  The wave function for the first excited states are shown in Fig.~\ref{fig-4-WaveFunc-excitation}. The first three excited states are all edge states. And we show the first excited state on the edge in Fig.~\ref{fig-4-WaveFunc-excitation}(a). Similar to the analysis of edge MZM and vortex MZM, our interest is also focused on the vortex states. Because the mini-gap is defined as the difference between the first vortex state and the vortex MZM, about $0.066\Delta_0$. Due to the spin property of this first vortex excitation, we see it is spin polarized down at vortex core, i.e., $u_\downarrow\neq0$ and $u_\uparrow=v_\uparrow=v_\downarrow=0$. It is consistent with the calculation in Ref\cite{mao_prb_2010,mzm_our_prl_2016}.

\subsection*{Appendix B : Normal Andreev Reflection}
  For normal Andreev reflection, incident electrons are reflected by SC device as holes with opposite spin direction.
  For comparison, we calculated the Andreev reflection coefficient in normal Andreev reflection case.
  The Andreev reflection coefficient $T_A$ here is defined by
  \begin{equation}
  T_A= \text{Tr} (r_{he}^{\dag} r_{he})
  \end{equation}

  \begin{figure}[!htbp]
  	\centering
  	\includegraphics[width=3.2in]{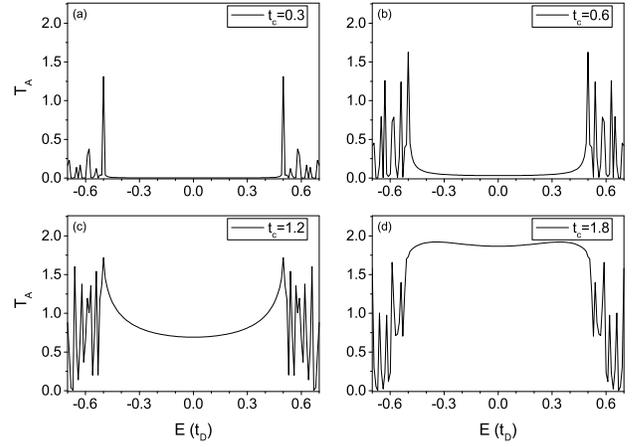}
  	\caption{\label{fig-6-NAR}The plots of normal Andreev reflection coefficient. The SC device is changed by a normal s-wave SC with parameters \( \Delta(\boldsymbol{r})= \Delta_0= 0.5 \), \(\alpha_R= 0\), magnetic field \(\boldsymbol{V}_{D}= 0\), \(\mu_D= 0.8 \), and \( \boldsymbol{V}_{L}= 0\), \(\mu_L= 0.9 \) for the spin unpolarized STM tip. Only the coupling coefficient \( t_c \) is shifted in (a)\(\sim\)(d).}
  \end{figure}

  As showed in Fig.~\ref{fig-6-NAR}, for tiny coupling coefficient \( t_c \), the Andreev reflection coefficient \( T_A \) vanished for all energy \( E \) lower than SC gap \( \Delta \). Since \( t_c \) increasing in certain range, the \( T_A \) that the reflection in SC gap contributed rises, and finally form a plateau valued \( 2 \) for two channels spin up and down. These behaves of Andreev reflection coefficient are consistent with the BTK theory.

\end{document}